\documentclass{vgtc}                          

\graphicspath{{figures/}{pictures/}{images/}{./}} 

\usepackage{times}                     

\usepackage{tabu}                      
\usepackage{booktabs}                  
\usepackage{lipsum}                    
\usepackage{mwe}                       

\usepackage{mathptmx}                  

\onlineid{0}

\vgtccategory{Research}

\vgtcinsertpkg

\title{From Graphs to Words: A Computer-Assisted Framework for the Production of Accessible Text Descriptions}

\author{Qiang Xu\thanks{e-mail: qiang.xu@polymtl.ca}
\and Thomas Hurtut\thanks{e-mail:thomas.hurtut@polymtl.ca}}
\affiliation{\scriptsize Polytechnique Montréal}

\teaser{
  \centering
  \includegraphics[width=\linewidth]{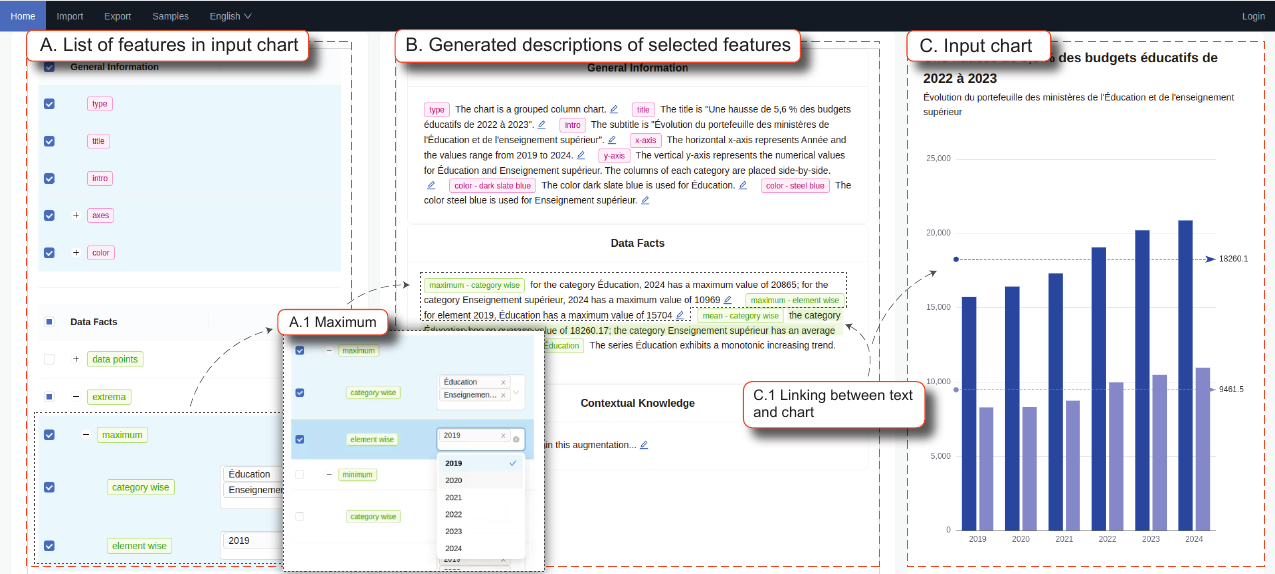}
  \caption{Main interface with three components: A. List of features in input chart, B. Generated descriptions of selected features, and C. Input chart.}
  \label{fig:teaser}
}

\abstract{
    In the digital landscape, the ubiquity of data visualizations in media underscores the necessity for accessibility to ensure inclusivity for all users, including those with visual impairments. Current visual content often fails to cater to the needs of screen reader users due to the absence of comprehensive textual descriptions. To address this gap, we propose in this paper a framework designed to empower media content creators to transform charts into descriptive narratives. This tool not only facilitates the understanding of complex visual data through text but also fosters a broader awareness of accessibility in digital content creation. Through the application of this framework, users can interpret and convey the insights of data visualizations more effectively, accommodating a diverse audience. Our evaluations reveal that this tool not only enhances the comprehension of data visualizations but also promotes new perspectives on the represented data, thereby broadening the interpretative possibilities for all users.
} 


\begin{document}

\firstsection{Introduction}

\maketitle

In our increasingly data-driven world, visualizations serve as pivotal tools for storytelling, offering clear and concise portrayals of complex datasets that cater to the swift comprehension needs of modern audiences \cite{5613452}. These visual representations transform abstract numbers into insightful narratives and intuitive graphics, making complex information more accessible to the general public. However, an exclusive reliance on these visual formats inherently excludes significant segments of the audience, including those with visual impairments. Extensive research has demonstrated that a substantial number of online visualizations are not compatible with screen-reader technologies, primarily due to the absence of adequate textual alternatives \cite{10.1145/3441852.3471202}. This limitation not only restricts the accessibility of such visualizations for visually impaired individuals and screen reader users but also affects those requiring auditory information for various activities, such as driving or multitasking. Consequently, to bridge this accessibility gap and ensure equitable access to information, the provision of textual descriptions for visualizations is not merely important—it is imperative.

The practice of providing detailed descriptions for non-textual elements on websites is gaining traction; however, the lack of precise and standardized guidelines significantly hampers consistency in these efforts \cite{10.1145/3457875}. The task of effectively describing visual elements in data-driven stories introduces multiple challenges. First, there is a notable learning curve associated with the skill of writing comprehensive descriptions for accessibility \cite{Matamala_Orero_2021}, as it requires an in-depth understanding of the critical details to be included. Second, data visualizations present complex datasets that demand advanced analytical skills \cite{6875906}. Creators must interpret visual patterns with precision, ensuring that the data is neither oversimplified nor misrepresented. Third, crafting such descriptions is often laborious and time-consuming, necessitating meticulous attention to detail. Despite diligent efforts, some descriptions may not adequately convey all the necessary details of the visual elements \cite{10.1145/2764916}. Recently, the machine learning community has explored deep learning-based methods to aid chart-text transformation \cite{9093592, obeid-hoque-2020-chart, kantharaj-etal-2022-chart, Cheng_2023_ICCV}. Yet, these black box models often lack transparency, challenging users' ability to fine-tune or fully comprehend the details in the generated outputs. In response to these issues, we propose the development of a novel framework designed to aid media content creators in interpreting charts and producing detailed textual descriptions. This preliminary research is tailored specifically for individuals who develop visualizations using Datawrapper\footnote{\url{https://www.datawrapper.de}}, a popular tool for those with basic proficiency in statistics and data analysis but without extensive programming knowledge.

Our framework makes three significant contributions to the field. First, we introduce a novel heuristic approach that affords authors greater control over the process of authoring descriptions, enhancing the accuracy and relevance of the output. Second, we have developed an automatic method for identifying features in visualizations, which streamlines different stages of the description process by highlighting critical data elements automatically. Third, we conducted a comprehensive user study to evaluate the practical utility of our framework and to identify challenges for future research. The study not only validates the effectiveness of our method but also highlights areas requiring further exploration to enhance the framework's applicability and functionality.

\begin{figure*}[t]
    \centering
    \includegraphics[width=\linewidth]{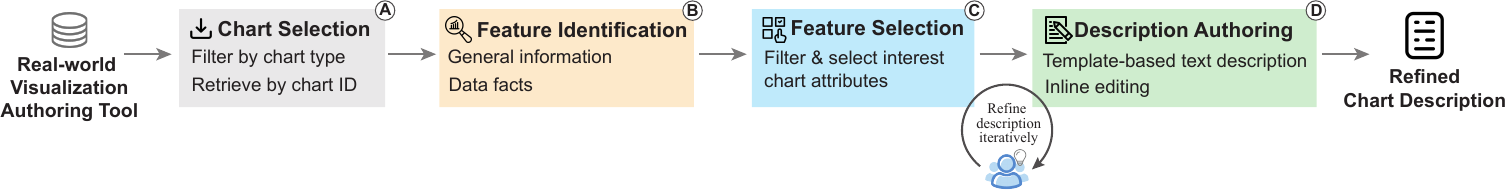}
    \caption{Workflow of the framework.}
    \label{fig:pipeline}
    \vspace{-4mm}
\end{figure*}

\section{Related Work}
The translation of visual data representations into accessible formats has received considerable attention across several fields, including accessibility, visualization, computer vision, and natural language processing.

The World Wide Web Consortium (W3C) plays a crucial role in setting guidelines and standards for web accessibility. One of its fundamental principles is ensuring that all non-textual web content provides a text alternative that fulfills the same function. The W3C particularly stresses the importance of crafting detailed descriptions for complex images such as charts, maps, and other visualizations. The process of converting these complex images into text involves a substantial reduction in information dimension, which brings forth critical considerations regarding the specific needs and preferences of the target audience. Consequently, recent studies increasingly aim to pinpoint what aspects of these descriptions are most valued by visually impaired people \cite{9552938, 9555469}. Concurrently, there is ongoing research exploring various design approaches for making visualizations more accessible \cite{10.1145/3544548.3581186}. This research includes developing innovative methods to navigate, comprehend, and interact with graphical data using non-visual interfaces like keyboards \cite{https://doi.org/10.1111/cgf.14519} and speech interfaces \cite{10.1145/3491102.3517431}.

Significant advancements in deep learning have been made towards automating the generation of text from visualizations and structured data. Chen et al. employed ResNet and LSTM architectures to generate natural language descriptions for figures in the FigCAP dataset \cite{9093592}, while Cheng et al. proposed a framework with transformer-based chart detection and pre-trained vision-language model \cite{Cheng_2023_ICCV}. Additionally, some researchers prioritize the underlying data rather than the visual representation of charts, focusing on data-to-text generation. Notable efforts in this area include the use of transformer-based models \cite{gong-etal-2019-enhanced, obeid-hoque-2020-chart} and encoder-decoder LSTM frameworks for analyzing time-series data \cite{10.1145/3399715.3399829}. In assessing the performance of these models, metrics like BLEU score \cite{10.3115/1073083.1073135} are commonly applied to test datasets. Despite their computational skills, deep learning models often encounter issues with generalization and credibility. Generalization problems occur when the models underperform with unfamiliar datasets, limiting their practical use. Credibility concerns stem from the models' opaque nature, which obscures the internal mechanisms and decision-making processes. To address these natural language generation issues, some research prototypes \cite{8805442, 10297564} have adopted template-based text generation methods, which offer more transparent approaches to creating textual descriptions from visual data.

\section{Design Challenges and Requirements}
To convert a chart to text, automatic methods that employ deep learning make it hard for users to integrate their contextual information. Moreover, these methodologies typically fall short in considering the precise content requirements necessary for meaningful descriptions, which also necessitate an accurate interpretation of the data. Such shortcomings prompt the need for alternative strategies. As a result, we decided to create a methodology that facilitates the exploration of data to refine automatically generated descriptions by adopting a heuristic method that incorporates a human-in-the-loop approach. Such an approach allows human expertise to improve the accuracy and relevance of the textual outputs but also ensures that the descriptions are appropriate and complete.

\subsection{Design Challenges}
We distill four design challenges from the literature review and discussions with experts in communication and accessibility.

\begin{itemize}
    \itemsep0em
    \item[C1.] \textbf{A design that does not fit well with existing systems and processes can cause difficulties for users to adopt it.} Users must invest time and effort to learn, which can lead to frustration and resistance to change.
    \item[C2.] \textbf{Properly describing a visualization requires a careful examination of all its features.} This requires a profound understanding of accessibility principles and the information needs of the target audience \cite{9552938}.
    \item[C3.] \textbf{The accuracy of the description depends on the correct interpretation.} Fully automated systems can make critical errors \cite{Salisbury_Kamar_Morris_2017} as their performance is tied to training data and may struggle with unfamiliar situations.
    \item[C4.] \textbf{Users may be resistant to fully trusting auto-generated outputs.} The transformation of format, i.e. chart to text, can make it difficult for users to discern how accurately the resulting text represents the original visual content \cite{7536106}.
\end{itemize}

\subsection{Design Requirements}
Based on the challenges, we derive four requirements to guide the design of our framework.
\begin{itemize}
    \itemsep0em
    \item[R1.] \textbf{The design should be compatible with the existing workflow.} Ensuring compatibility with tools commonly used by media content creators is crucial, as it minimizes the learning curve and reduces operational disruptions.
    \item[R2.] \textbf{The tool should extract features from the input chart and present them in a structured format.} These features should be identified and verified through a heuristic approach. A structured presentation helps users categorize complex information, thereby enhancing the efficiency of data analysis.
    \item[R3.] \textbf{Each feature needs a corresponding description.} To facilitate the refinement process, generated descriptions must be accurate, editable, and rearrangeable.
    \item[R4.] \textbf{Generated descriptions should be connected to the original chart using linking and brushing techniques.} To foster trust, the design should incorporate visual cues to highlight key points, guide the viewer's focus, and illustrate the descriptions.
\end{itemize}

\section{Design and Implementation}
Our design and implementation process is guided by literature reviews from visualization and accessibility research. Throughout the implementation, we collaborated with experts to validate and refine our design.

\subsection{Workflow}
The workflow of our framework is designed to streamline the conversion of an input chart into a textual description. Initially, users input a chart (\cref{fig:pipeline}A), prompting the framework to extract both data and metadata and to identify key features essential for detailed analysis (\cref{fig:pipeline}B). This marks the beginning of an interactive phase where users play an important role. In this phase, users critically assess the extracted features, select those most relevant, and meticulously refine the descriptions for each (\cref{fig:pipeline}C and \cref{fig:pipeline}D). This cycle of selection and refinement is iteratively repeated as necessary to enhance the narrative and accuracy of the data’s story.

\subsection{Chart Selection}
In the initial phase, users are required to give an input chart. The framework is designed to integrate with the Datawrapper database (\textbf{R1}), enabling users to filter charts by type and retrieve them using their ID. The chart selection page displays charts from the database in descending order by creation date, with the most recent charts on the first page. Each chart is represented by a thumbnail, and clicking on a thumbnail activates a preview on the right side of the page. This ensures that users can quickly and easily access the specific data visualization they need.

\subsection{Feature Identification and Selection}
The identified features are displayed in an interactive list (\cref{fig:teaser}A), facilitating a comprehensive understanding of the elements comprising the input chart (\textbf{R2}). The process of feature identification and selection contains three distinct phases: \textbf{1) data extraction}, where relevant data are retrieved from the input; \textbf{2) feature detection}, during which the system identifies key elements necessary for analysis; and \textbf{3) feature presentation and selection}, where these elements are displayed for user interaction and selection. 

\textbf{Data Extraction.} 
Given an input chart, the framework extracts raw data and metadata by connecting to the database using APIs. The raw data, presented in tabular format, refer to the foundational elements that construct the chart, such as numerical and categorical values. Metadata comprises a variety of details, including the chart's title, type, and any accompanying notes. Additionally, it captures aesthetic aspects like color schemes and functional attributes such as whether the data is sorted and its sorting order.

\textbf{Feature Detection.}
The framework conducts an automatic analysis of the extracted data and metadata to identify key features. The detection is based on chart type and currently supports all basic statistical charts and their variants, such as bar charts (split bars, stacked bars, grouped bars, etc.), area charts, line charts, and pie charts. For each type of chart, the framework detects key features and classifies them into the following categories: general information and data facts \cite{9555469}.

For general information, the framework captures important details such as chart type, main title, subtitles, footnotes, axes, and color schemes. The comprehensive understanding of visualization goes beyond mere data representation; it necessitates an elaborate description of all graphical elements. These detailed descriptions are crucial for visually impaired people, as they enable them to form a mental representation of the visualization, enhancing their comprehension of the overall design \cite{9552938}.

Regarding data facts, the framework analyzes the numerical and statistical elements of the chart. It identifies specific data points and calculates statistical measures such as extrema, mean, standard deviation, and median. Additionally, it recognizes outliers—data points that markedly differ from the main observations. When the chart involves numerical data on the independent axis, the framework assesses correlations and trends. These facts are derived from the low-level analytical tasks that viewers typically perform while engaging with a visualization \cite{10.1109/INFOVIS.2005.24}.

In addition to those two categories, we have also incorporated a section for contextual knowledge, which allows users to input information explaining the purpose behind including the chart.

\textbf{Feature Presentation and Selection.}
The detected features are displayed as checkbox lists, as illustrated in \cref{fig:teaser}A. To enhance user understanding, elements within each category are color-coded to signify their semantic significance—for instance, pink for general information and green for data facts. For visualizations such as bar and column charts that exhibit univariate data, the features are presented in a straightforward series of checkboxes. Conversely, for more complex, multivariate visualizations like grouped column charts, the interface incorporates an additional dropdown list (\cref{fig:teaser}A.1). This feature allows users to select specific variables, thereby accommodating the intricacies of multivariate data analysis.

\subsection{Description Authoring}
Upon the selection of a checkbox, the framework initiates template-based text generation to fill the description component (\textbf{R3}) as shown in \cref{fig:teaser}B. Each segment of the generated description is associated with the feature's name and is color-coordinated with the corresponding checkbox for clarity. The framework employs heuristic analysis of the features within the input chart, where each specific condition is connected to a designated text template. In charts depicting univariate data, selecting a checkbox automatically generates a preset textual description. Conversely, in charts with multivariate data, the text is dynamically updated based on the selected variables, enabling detailed comparisons both within and between groups \cite{9552208}. This functionality ensures that the displayed information is tailored to the specific categories, offering personalized and pertinent descriptions that align with the user’s comprehension and interaction with the data.

To ensure that the generated text meets high standards of clarity and stylistic quality, the framework allows users to fine-tune descriptions. Users can edit or rearrange text segments effortlessly by dragging the associated tags. Furthermore, when users hover their cursor over any part of the description, an animation activates on the corresponding section of the input chart (\cref{fig:teaser}C). This animation (\textbf{R4}), combined with visual cues, emphasizes the relevance of the text, offering a dynamic and interactive method for illustrating the relationship between textual descriptions and visual data \cite{10.5555/3290776.3290796, 8440860}.

\subsection{Implementation}
The framework\footnote{Source code: \url{https://github.com/yuri7718/graphs2words}} is developed as a web-based application utilizing React and Flask\footnote{\url{https://react.dev}, \url{https://flask.palletsprojects.com}}. To initiate the use of a chart within the framework, users must first generate a token in their Datawrapper account. This token serves to authenticate the user, granting access to and enabling the retrieval of charts specific to their account.

\textbf{Chart Data Extraction.} 
For feature identification, the backend retrieves data using Datawrapper APIs. Specifically, it fetches three types of information: 1) metadata in JSON format containing title, type, and other configuration information; 2) raw data; and 3) chart in SVG format. Colors are extracted from the SVG to reconstruct the visualization using the raw data and metadata details, such as chart type.

\textbf{Chart Color Identification.}
To handle color identification, we used the list of CSS3 extended color names which consist of key-value pairs of hex color codes and the corresponding names. All RGB color codes, including those extracted from SVG, are converted to the CIELAB \cite{Luo2014} color space. This transformation allows us to use the nearest neighbor algorithm \cite{9065747}, as CIELAB aims to approximate human vision rather than just mimic the physical attributes of colors. This ensures that the closest color name can be effectively identified for each detected color in the input chart.

\textbf{Data Facts.}
Data facts such as extrema, mean, standard deviation, and median are calculated from raw data. To detect outliers, we employed the Interquartile Range method \cite{vinutha2018detection}. This statistical technique is designed to identify anomalies by defining acceptable boundaries based on the spread of data. To analyze numerical data exhibiting a trend, our initial step is to determine whether the data exhibits a monotonic behavior. If not, we segment the data into intervals where values are either rising, falling, or remaining constant. Should the total number of these intervals exceed a specific threshold, it indicates considerable variability in the trend. In such cases, we further refine our analysis by calculating the slope for each interval, allowing us to identify and highlight those intervals that demonstrate significant changes with respect to the independent axis, thereby focusing on the most critical variations within the data.

\section{Evaluation}
We conducted a user study\footnote{User study approved by Research Ethics Board of Polytechnique Montréal (ID CER-2324-51-D).} involving three participants (\textbf{P1-P3}), all of whom specialize in media content creation. The participants are experienced journalists with an average of 10 years of professional experience and prior experience using Datawrapper for visualization purposes. Each participant was presented with the tool and given the opportunity to explore its features independently. They could choose to experiment with charts from the provided samples or import a new chart of their choice. The samples page contained two datasets, one with 401 charts and the other with 1035 charts, offering a variety of chart types based on tabular data. Following this exploratory phase, we conducted individual semi-structured interviews lasting approximately 30 minutes. During these interviews, we posed a series of questions to assess whether our design met the established requirements.

\textbf{R1. The design should be compatible with the existing workflow.} Feedback from participants emphasizes the design's ease of use, suggesting that it aligns well with their existing workflow and does not introduce unnecessary complexity. In general, users value the effectiveness and simplicity of the main interface. As \textbf{P3} remarked: \textit{``The tool is simple and intuitive. I didn't feel that I had to click everywhere to understand how to use it. Usually, you click and see, and it works. I found it very easy to use.''}

\textbf{R2. The tool should extract features from the input chart and present them in a structured format.} 
Some participants appreciated the volume of information, while others felt it could be overwhelming. The design handles this by providing options for filtering out unnecessary details. \textbf{P2} expressed appreciation for the extensive information, stating: \textit{``I liked the fact that the list is exhaustive. There are a lot of elements, and I can choose the most relevant ones."} \textbf{P3} commented on the feature richness, adding: \textit{``There is pretty much everything. There is a lot of information, but it's still easy to hide things you don't need. You just have to get into the habit of unchecking certain things so that it's less intense. It's still easy to manage the amount of information."}

\textbf{R3. Each feature needs a corresponding description.} 
Participants found the descriptions helpful in understanding the charts. \textbf{P1} appreciated the ability to identify significant moments, explaining: \textit{``Being able to spot significant moments is remarkable. This brings out interesting points and reminds us where the increases and decreases are. For someone who only has these descriptions to understand the chart, it gives a basis."} \textbf{P3} liked the provided statistical values, saying: \textit{``The fact that it gives the max, min, and average is interesting and very useful because that’s often what we try to tell."} Regarding interactions with the text, \textbf{P2} and \textbf{P3} enjoyed the ability to rearrange the descriptions to create a narrative. 

\textbf{R4. Generated descriptions should be connected to the original chart using linking and brushing techniques.}
Feedback from participants indicates a positive reception towards the use of animated visual cues. \textbf{P1} found the animation interesting, noting that it allows users to contextualize and tell the story by emphasizing key points. \textbf{P2} thought the annotations were very helpful for clearly illustrating what the descriptions explain and describe.

Some participants pointed out that the framework not only helps users determine what to write but also facilitates graph reinterpretation for deeper insights. \textbf{P2} highlighted that it often allows for a new understanding. \textbf{P3} elaborated on its utility, explaining that it could also be used for analysis. Users may find it useful to see specific information without needing to go to a spreadsheet for calculations, as searching for information in a spreadsheet can be more challenging. With the framework, relevant details are readily available visually with just a click, making it simple and efficient.

\section{Limitations and Future Work}
\textbf{Limitations.} Regarding the framework's efficiency in saving time, although \textbf{P2} considers it a primary strength, other participants noted that the amount of information could potentially overwhelm average users, particularly those with limited data understanding. The tool offers extensive details about the underlying data but falls short in suggesting high-level communicative objectives, such as presenting the narrative or the context that the data encapsulated. Furthermore, visualizations serve dual roles in data presentation, being both communicative and explorative \cite{6064988}. The framework relies on users to generate descriptions, inherently reflecting their unique perspectives. Consequently, integrating the explorative nature of data remains a significant challenge, as it requires a delicate balance between detailed analysis and clear, concise storytelling.

\textbf{Future work.} A common suggestion from the feedback was to make output descriptions less technical and more narrative-driven, emphasizing general trends over detailed data points. This could be achieved by improving template wording using LLMs and quantifying changes in data trends to offer users a more nuanced understanding of magnitude. Currently, the framework lists every feature of an input chart. To simplify it for general users, we can improve it by automatically recommending top insights, with these checkboxes pre-selected by default. Additionally, ongoing research into accessibility is crucial, as preferences for description length and detail vary widely \cite{10.1145/3313831.3376404}, and no universal standard exists yet for optimizing these elements in our user interface.

\acknowledgments{
This research was funded by an NSERC Alliance Grant ALLRP-561132-20, in collaboration with Le Devoir, and an IVADO Grant in collaboration with Radio-Canada. We acknowledge and appreciate the help and contribution of Le Devoir and Radio-Canada as well as all the participants in different stages of this study.
}

\bibliographystyle{abbrv-doi}

\bibliography{template}

\begin{thebibliography}{10}

\bibitem{10.1109/INFOVIS.2005.24}
R.~Amar, J.~Eagan, and J.~Stasko.
\newblock Low-level components of analytic activity in information visualization.
\newblock In {\em Proceedings of the IEEE Symposium on Information Visualization}, p.~15. IEEE Computer Society, USA, 2005. doi: {{%
10\hspace{.1pt}\discretionary{.}{%
}{.}\hspace{.4pt}1109\discretionary{/}{%
}{/}INFOVIS\hspace{.1pt}\discretionary{.}{%
}{.}\hspace{.4pt}2005\hspace{.1pt}\discretionary{.}{%
}{.}\hspace{.4pt}24}}


\bibitem{6875906}
J.~Boy, R.~A. Rensink, E.~Bertini, and J.-D. Fekete.
\newblock A principled way of assessing visualization literacy.
\newblock {\em IEEE Transactions on Visualization and Computer Graphics}, 20(12):1963--1972, 2014. doi: {{%
10\hspace{.1pt}\discretionary{.}{%
}{.}\hspace{.4pt}1109\discretionary{/}{%
}{/}TVCG\hspace{.1pt}\discretionary{.}{%
}{.}\hspace{.4pt}2014\hspace{.1pt}\discretionary{.}{%
}{.}\hspace{.4pt}2346984}}


\bibitem{9093592}
C.~Chen, R.~Zhang, E.~Koh, S.~Kim, S.~Cohen, and R.~Rossi.
\newblock Figure captioning with relation maps for reasoning.
\newblock In {\em Proceedings of IEEE Winter Conference on Applications of Computer Vision}, pp. 1526--1534, 2020. doi: {{%
10\hspace{.1pt}\discretionary{.}{%
}{.}\hspace{.4pt}1109\discretionary{/}{%
}{/}WACV45572\hspace{.1pt}\discretionary{.}{%
}{.}\hspace{.4pt}2020\hspace{.1pt}\discretionary{.}{%
}{.}\hspace{.4pt}9093592}}


\bibitem{Cheng_2023_ICCV}
Z.-Q. Cheng, Q.~Dai, and A.~G. Hauptmann.
\newblock Chart{R}eader: A unified framework for chart derendering and comprehension without heuristic rules.
\newblock In {\em Proceedings of the IEEE Conference on Computer Vision}, pp. 22202--22213, Oct 2023.

\bibitem{7536106}
A.~Dasgupta, J.-Y. Lee, R.~Wilson, R.~A. Lafrance, N.~Cramer, K.~Cook, and S.~Payne.
\newblock Familiarity vs trust: A comparative study of domain scientists' trust in visual analytics and conventional analysis methods.
\newblock {\em IEEE Transactions on Visualization and Computer Graphics}, 23(1):271--280, 2017. doi: {{%
10\hspace{.1pt}\discretionary{.}{%
}{.}\hspace{.4pt}1109\discretionary{/}{%
}{/}TVCG\hspace{.1pt}\discretionary{.}{%
}{.}\hspace{.4pt}2016\hspace{.1pt}\discretionary{.}{%
}{.}\hspace{.4pt}2598544}}


\bibitem{gong-etal-2019-enhanced}
L.~Gong, J.~Crego, and J.~Senellart.
\newblock Enhanced transformer model for data-to-text generation.
\newblock In {\em Proceedings of the Workshop on Neural Generation and Translation}, pp. 148--156. Association for Computational Linguistics, China, Nov 2019. doi: {{%
10\hspace{.1pt}\discretionary{.}{%
}{.}\hspace{.4pt}18653\discretionary{/}{%
}{/}v1\discretionary{/}{%
}{/}D19\discretionary{%
}{-}{-}5615}}


\bibitem{6064988}
J.~Hullman and N.~Diakopoulos.
\newblock Visualization rhetoric: Framing effects in narrative visualization.
\newblock {\em IEEE Transactions on Visualization and Computer Graphics}, 17(12):2231--2240, 2011. doi: {{%
10\hspace{.1pt}\discretionary{.}{%
}{.}\hspace{.4pt}1109\discretionary{/}{%
}{/}TVCG\hspace{.1pt}\discretionary{.}{%
}{.}\hspace{.4pt}2011\hspace{.1pt}\discretionary{.}{%
}{.}\hspace{.4pt}255}}


\bibitem{9552938}
C.~Jung, S.~Mehta, A.~Kulkarni, Y.~Zhao, and Y.-S. Kim.
\newblock Communicating visualizations without visuals: Investigation of visualization alternative text for people with visual impairments.
\newblock {\em IEEE Transactions on Visualization and Computer Graphics}, 28(1):1095--1105, 2022. doi: {{%
10\hspace{.1pt}\discretionary{.}{%
}{.}\hspace{.4pt}1109\discretionary{/}{%
}{/}TVCG\hspace{.1pt}\discretionary{.}{%
}{.}\hspace{.4pt}2021\hspace{.1pt}\discretionary{.}{%
}{.}\hspace{.4pt}3114846}}


\bibitem{kantharaj-etal-2022-chart}
S.~Kantharaj, R.~T. Leong, X.~Lin, A.~Masry, M.~Thakkar, E.~Hoque, and S.~Joty.
\newblock Chart-to-text: A large-scale benchmark for chart summarization.
\newblock In {\em Proceedings of the Association for Computational Linguistics}, pp. 4005--4023. Association for Computational Linguistics, Ireland, may 2022. doi: {{%
10\hspace{.1pt}\discretionary{.}{%
}{.}\hspace{.4pt}18653\discretionary{/}{%
}{/}v1\discretionary{/}{%
}{/}2022\hspace{.1pt}\discretionary{.}{%
}{.}\hspace{.4pt}acl\discretionary{%
}{-}{-}long\hspace{.1pt}\discretionary{.}{%
}{.}\hspace{.4pt}277}}


\bibitem{10.5555/3290776.3290796}
S.~Latif, D.~Liu, and F.~Beck.
\newblock Exploring interactive linking between text and visualization.
\newblock In {\em Proceedings of the Eurographics/IEEE Conference on Visualization}, p. 91–94. Eurographics Association, Goslar, DEU, 2018.

\bibitem{10297564}
F.~Lei, Y.~Ma, A.~S. Fotheringham, E.~A. Mack, Z.~Li, M.~Sachdeva, S.~Bardin, and R.~Maciejewski.
\newblock Geo{E}xplainer: A visual analytics framework for spatial modeling contextualization and report generation.
\newblock {\em IEEE Transactions on Visualization and Computer Graphics}, 30(1):1391--1401, 2024. doi: {{%
10\hspace{.1pt}\discretionary{.}{%
}{.}\hspace{.4pt}1109\discretionary{/}{%
}{/}TVCG\hspace{.1pt}\discretionary{.}{%
}{.}\hspace{.4pt}2023\hspace{.1pt}\discretionary{.}{%
}{.}\hspace{.4pt}3327359}}


\bibitem{9555469}
A.~Lundgard and A.~Satyanarayan.
\newblock Accessible visualization via natural language descriptions: A four-level model of semantic content.
\newblock {\em IEEE Transactions on Visualization and Computer Graphics}, 28(1):1073--1083, 2022. doi: {{%
10\hspace{.1pt}\discretionary{.}{%
}{.}\hspace{.4pt}1109\discretionary{/}{%
}{/}TVCG\hspace{.1pt}\discretionary{.}{%
}{.}\hspace{.4pt}2021\hspace{.1pt}\discretionary{.}{%
}{.}\hspace{.4pt}3114770}}


\bibitem{Luo2014}
M.~R. Luo.
\newblock {\em CIELAB}, pp. 1--7.
\newblock Springer Berlin Heidelberg, Berlin, Heidelberg, 2014. doi: {{%
10\hspace{.1pt}\discretionary{.}{%
}{.}\hspace{.4pt}1007\discretionary{/}{%
}{/}978\discretionary{%
}{-}{-}3\discretionary{%
}{-}{-}642\discretionary{%
}{-}{-}27851\discretionary{%
}{-}{-}8\_11\discretionary{%
}{-}{-}1}}


\bibitem{10.1145/3457875}
K.~Marriott, B.~Lee, M.~Butler, E.~Cutrell, K.~Ellis, C.~Goncu, M.~Hearst, K.~McCoy, and D.~A. Szafir.
\newblock Inclusive data visualization for people with disabilities: a call to action.
\newblock {\em Interactions}, 28(3):47–51, apr 2021. doi: {{%
10\hspace{.1pt}\discretionary{.}{%
}{.}\hspace{.4pt}1145\discretionary{/}{%
}{/}3457875}}


\bibitem{Matamala_Orero_2021}
A.~Matamala and P.~Orero.
\newblock Designing a course on audio description and defining the main competences of the future professional.
\newblock {\em Linguistica Antverpiensia, New Series – Themes in Translation Studies}, 6, Oct. 2021. doi: {{%
10\hspace{.1pt}\discretionary{.}{%
}{.}\hspace{.4pt}52034\discretionary{/}{%
}{/}lanstts\hspace{.1pt}\discretionary{.}{%
}{.}\hspace{.4pt}v6i\hspace{.1pt}\discretionary{.}{%
}{.}\hspace{.4pt}195}}


\bibitem{10.1145/2764916}
V.~S. Morash, Y.-T. Siu, J.~A. Miele, L.~Hasty, and S.~Landau.
\newblock Guiding novice web workers in making image descriptions using templates.
\newblock {\em ACM Transactions on Accessible Computing}, 7(4), nov 2015. doi: {{%
10\hspace{.1pt}\discretionary{.}{%
}{.}\hspace{.4pt}1145\discretionary{/}{%
}{/}2764916}}


\bibitem{obeid-hoque-2020-chart}
J.~Obeid and E.~Hoque.
\newblock Chart-to-text: Generating natural language descriptions for charts by adapting the transformer model.
\newblock In {\em Proceedings of the International Conference on Natural Language Generation}, pp. 138--147. Association for Computational Linguistics, Ireland, dec 2020. doi: {{%
10\hspace{.1pt}\discretionary{.}{%
}{.}\hspace{.4pt}18653\discretionary{/}{%
}{/}v1\discretionary{/}{%
}{/}2020\hspace{.1pt}\discretionary{.}{%
}{.}\hspace{.4pt}inlg\discretionary{%
}{-}{-}1\hspace{.1pt}\discretionary{.}{%
}{.}\hspace{.4pt}20}}


\bibitem{10.3115/1073083.1073135}
K.~Papineni, S.~Roukos, T.~Ward, and W.-J. Zhu.
\newblock {BLEU}: a method for automatic evaluation of machine translation.
\newblock In {\em Proceedings of the Annual Meeting on Association for Computational Linguistics}, p. 311–318. Association for Computational Linguistics, USA, 2002. doi: {{%
10\hspace{.1pt}\discretionary{.}{%
}{.}\hspace{.4pt}3115\discretionary{/}{%
}{/}1073083\hspace{.1pt}\discretionary{.}{%
}{.}\hspace{.4pt}1073135}}


\bibitem{Salisbury_Kamar_Morris_2017}
E.~Salisbury, E.~Kamar, and M.~Morris.
\newblock Toward scalable social alt text: Conversational crowdsourcing as a tool for refining vision-to-language technology for the blind.
\newblock {\em Proceedings of the AAAI Conference on Human Computation and Crowdsourcing}, 5(1):147--156, Sep. 2017. doi: {{%
10\hspace{.1pt}\discretionary{.}{%
}{.}\hspace{.4pt}1609\discretionary{/}{%
}{/}hcomp\hspace{.1pt}\discretionary{.}{%
}{.}\hspace{.4pt}v5i1\hspace{.1pt}\discretionary{.}{%
}{.}\hspace{.4pt}13301}}


\bibitem{5613452}
E.~Segel and J.~Heer.
\newblock Narrative visualization: Telling stories with data.
\newblock {\em IEEE Transactions on Visualization and Computer Graphics}, 16(6):1139--1148, 2010. doi: {{%
10\hspace{.1pt}\discretionary{.}{%
}{.}\hspace{.4pt}1109\discretionary{/}{%
}{/}TVCG\hspace{.1pt}\discretionary{.}{%
}{.}\hspace{.4pt}2010\hspace{.1pt}\discretionary{.}{%
}{.}\hspace{.4pt}179}}


\bibitem{10.1145/3441852.3471202}
A.~Sharif, S.~S. Chintalapati, J.~O. Wobbrock, and K.~Reinecke.
\newblock Understanding screen-reader users’ experiences with online data visualizations.
\newblock In {\em Proceedings of the ACM Conference on Computers and Accessibility}. Association for Computing Machinery, USA, 2021. doi: {{%
10\hspace{.1pt}\discretionary{.}{%
}{.}\hspace{.4pt}1145\discretionary{/}{%
}{/}3441852\hspace{.1pt}\discretionary{.}{%
}{.}\hspace{.4pt}3471202}}


\bibitem{10.1145/3491102.3517431}
A.~Sharif, O.~H. Wang, A.~T. Muongchan, K.~Reinecke, and J.~O. Wobbrock.
\newblock Vox{L}ens: Making online data visualizations accessible with an interactive javascript plug-in.
\newblock In {\em Proceedings of the ACM Conference on Human Factors in Computing Systems}. Association for Computing Machinery, USA, 2022. doi: {{%
10\hspace{.1pt}\discretionary{.}{%
}{.}\hspace{.4pt}1145\discretionary{/}{%
}{/}3491102\hspace{.1pt}\discretionary{.}{%
}{.}\hspace{.4pt}3517431}}


\bibitem{10.1145/3399715.3399829}
A.~Spreafico and G.~Carenini.
\newblock Neural data-driven captioning of time-series line charts.
\newblock In {\em Proceedings of the International Conference on Advanced Visual Interfaces}. Association for Computing Machinery, New York, NY, USA, 2020. doi: {{%
10\hspace{.1pt}\discretionary{.}{%
}{.}\hspace{.4pt}1145\discretionary{/}{%
}{/}3399715\hspace{.1pt}\discretionary{.}{%
}{.}\hspace{.4pt}3399829}}


\bibitem{8440860}
A.~Srinivasan, S.~M. Drucker, A.~Endert, and J.~Stasko.
\newblock Augmenting visualizations with interactive data facts to facilitate interpretation and communication.
\newblock {\em IEEE Transactions on Visualization and Computer Graphics}, 25(1):672--681, 2019. doi: {{%
10\hspace{.1pt}\discretionary{.}{%
}{.}\hspace{.4pt}1109\discretionary{/}{%
}{/}TVCG\hspace{.1pt}\discretionary{.}{%
}{.}\hspace{.4pt}2018\hspace{.1pt}\discretionary{.}{%
}{.}\hspace{.4pt}2865145}}


\bibitem{10.1145/3313831.3376404}
A.~Stangl, M.~R. Morris, and D.~Gurari.
\newblock "{P}erson, shoes, tree. is the person naked?" what people with vision impairments want in image descriptions.
\newblock In {\em Proceedings of the ACM Conference on Human Factors in Computing Systems}, p. 1–13. Association for Computing Machinery, New York, NY, USA, 2020. doi: {{%
10\hspace{.1pt}\discretionary{.}{%
}{.}\hspace{.4pt}1145\discretionary{/}{%
}{/}3313831\hspace{.1pt}\discretionary{.}{%
}{.}\hspace{.4pt}3376404}}


\bibitem{9065747}
K.~Taunk, S.~De, S.~Verma, and A.~Swetapadma.
\newblock A brief review of nearest neighbor algorithm for learning and classification.
\newblock In {\em International Conference on Intelligent Computing and Control Systems}, pp. 1255--1260, 2019. doi: {{%
10\hspace{.1pt}\discretionary{.}{%
}{.}\hspace{.4pt}1109\discretionary{/}{%
}{/}ICCS45141\hspace{.1pt}\discretionary{.}{%
}{.}\hspace{.4pt}2019\hspace{.1pt}\discretionary{.}{%
}{.}\hspace{.4pt}9065747}}


\bibitem{10.1145/3544548.3581186}
J.~R. Thompson, J.~J. Martinez, A.~Sarikaya, E.~Cutrell, and B.~Lee.
\newblock Chart reader: Accessible visualization experiences designed with screen reader users.
\newblock In {\em Proceedings of the ACM Conference on Human Factors in Computing Systems}. Association for Computing Machinery, USA, 2023. doi: {{%
10\hspace{.1pt}\discretionary{.}{%
}{.}\hspace{.4pt}1145\discretionary{/}{%
}{/}3544548\hspace{.1pt}\discretionary{.}{%
}{.}\hspace{.4pt}3581186}}


\bibitem{vinutha2018detection}
H.~Vinutha, B.~Poornima, and B.~Sagar.
\newblock Detection of outliers using interquartile range technique from intrusion dataset.
\newblock In {\em Proceedings of the International Conference on Frontiers of Intelligent Computing}, pp. 511--518. Springer, 2018.

\bibitem{8805442}
Y.~Wang, Z.~Sun, H.~Zhang, W.~Cui, K.~Xu, X.~Ma, and D.~Zhang.
\newblock Data{S}hot: Automatic generation of fact sheets from tabular data.
\newblock {\em IEEE Transactions on Visualization and Computer Graphics}, 26(1):895--905, 2020. doi: {{%
10\hspace{.1pt}\discretionary{.}{%
}{.}\hspace{.4pt}1109\discretionary{/}{%
}{/}TVCG\hspace{.1pt}\discretionary{.}{%
}{.}\hspace{.4pt}2019\hspace{.1pt}\discretionary{.}{%
}{.}\hspace{.4pt}2934398}}


\bibitem{9552208}
C.~Xiong, V.~Setlur, B.~Bach, E.~Koh, K.~Lin, and S.~Franconeri.
\newblock Visual arrangements of bar charts influence comparisons in viewer takeaways.
\newblock {\em IEEE Transactions on Visualization and Computer Graphics}, 28(1):955--965, 2022. doi: {{%
10\hspace{.1pt}\discretionary{.}{%
}{.}\hspace{.4pt}1109\discretionary{/}{%
}{/}TVCG\hspace{.1pt}\discretionary{.}{%
}{.}\hspace{.4pt}2021\hspace{.1pt}\discretionary{.}{%
}{.}\hspace{.4pt}3114823}}


\bibitem{https://doi.org/10.1111/cgf.14519}
J.~Zong, C.~Lee, A.~Lundgard, J.~Jang, D.~Hajas, and A.~Satyanarayan.
\newblock Rich screen reader experiences for accessible data visualization.
\newblock {\em Computer Graphics Forum}, 41(3):15--27, 2022. doi: {{%
10\hspace{.1pt}\discretionary{.}{%
}{.}\hspace{.4pt}1111\discretionary{/}{%
}{/}cgf\hspace{.1pt}\discretionary{.}{%
}{.}\hspace{.4pt}14519}}


\end{thebibliography}
\end{document}